\documentclass{PoS}
\usepackage{amsmath,amssymb,fleqn,url,color,here,graphics,graphicx,multirow,wrapfig}
\usepackage{etex}
\usepackage{pbox}
\usepackage{bigstrut}
\usepackage{soul,xcolor}
\usepackage[ separate-uncertainty  = true,
   uncertainty-separator =  {\,},
   mode = text,
   exponent-product = \cdot,
   output-product = \cdot,
   output-decimal-marker ={.},
   multi-part-units      =  single,
   range-phrase          = {--}]{siunitx}
\newcommand{\slim}{\mskip 1.5mu}              

\newcommand{\phiH}{\phi _h}
\newcommand{\phiS}{\phi _S}
\newcommand{\gvc}{GeV/$c$ }

\newcommand{\red}{\textcolor[rgb]{1.00,0.00,0.00}}
\newcommand{\blue}{\textcolor[rgb]{0.00,0.00,1.00}}
\newcommand{\green}{\textcolor[rgb]{0.00,0.59,0.00}}

\setstcolor{red}
\title{Measurement of longitudinal-target-polarization dependent azimuthal asymmetries in SIDIS at COMPASS experiment}

\ShortTitle{Measurement of LSAs in SIDIS at COMPASS experiment}

\author{\speaker{Bakur Parsamyan}\\
        CERN, University of Turin and INFN section of Turin\\
        E-mail: \email{bakur.parsamyan@cern.ch}}

\abstract{Preliminary results obtained by the COMPASS experiment for longitudinal-spin-dependent azimuthal asymmetries in single-hadron muoproduction off protons are presented for the first time.
The analysis was carried out on the full data sample collected by COMPASS with longitudinally polarized proton ($NH_3$) target in 2007 and 2011 using 160~\gvc and 200~\gvc muon beams, correspondingly.
Within QCD parton model approach extracted asymmetries are giving access to the specific convolutions of certain transverse-momentum-dependent \textit{twist-2} and \textit{higher twist} parton distribution functions and fragmentation functions. In order to access different features of the involved distributions, the asymmetries are extracted as functions of different kinematic variables for both positive and negative single-hadron productions. The significant amount of collected data and wide kinematic range give unique importance to the results.
Obtained asymmetries are compared with previous measurements performed by HERMES experiment and with available model predictions.}

\FullConference{XXV International Workshop on Deep-Inelastic Scattering and Related Subjects\\
		3-7 April 2017\\
		University of Birmingham, UK}

\begin{document}
\section{Introduction}
\label{sec:intro}

Within QCD, quark transverse-momentum-dependent (TMD) parton distribution functions (PDFs) play an important role in the theoretical description of high energy reactions providing a three-dimensional picture of a fast moving nucleon in momentum space. Recent decades were marked by strong interest and significant progress in both theoretical and experimental studies of spin-(in)dependent TMD PDFs, for recent reviews see Refs.~\cite{Perdekamp:2015vwa,Boglione:2015zyc,Aidala:2012mv}. In QCD parton model approach at leading order (LO) the spin and quark-transverse-momentum structure of the nucleon is fully described by eight \textit{twist-2} TMD PDFs~\cite{Kotzinian:1994dv,Bacchetta:2006tn}. Going beyond the leading order approximation, sixteen \textit{twist-3} PDFs related to quark-gluon correlations have to be introduced already at the first sub-leading accuracy level~\cite{Bacchetta:2006tn}.
The whole set of \textit{leading-} and \textit{subleading-twist} TMD PDFs of the nucleon can be accessed through measurement of the spin (in)dependent azimuthal asymmetries, arising in semi-inclusive hadron production in deep-inelastic lepton-nucleon scattering, $\ell \,N \rightarrow \ell^\prime \,h \, X$ (hereafter referred to as SIDIS).

\begin{wrapfigure}{r}{6.0cm}
\centering
\includegraphics[width=0.39\textwidth]{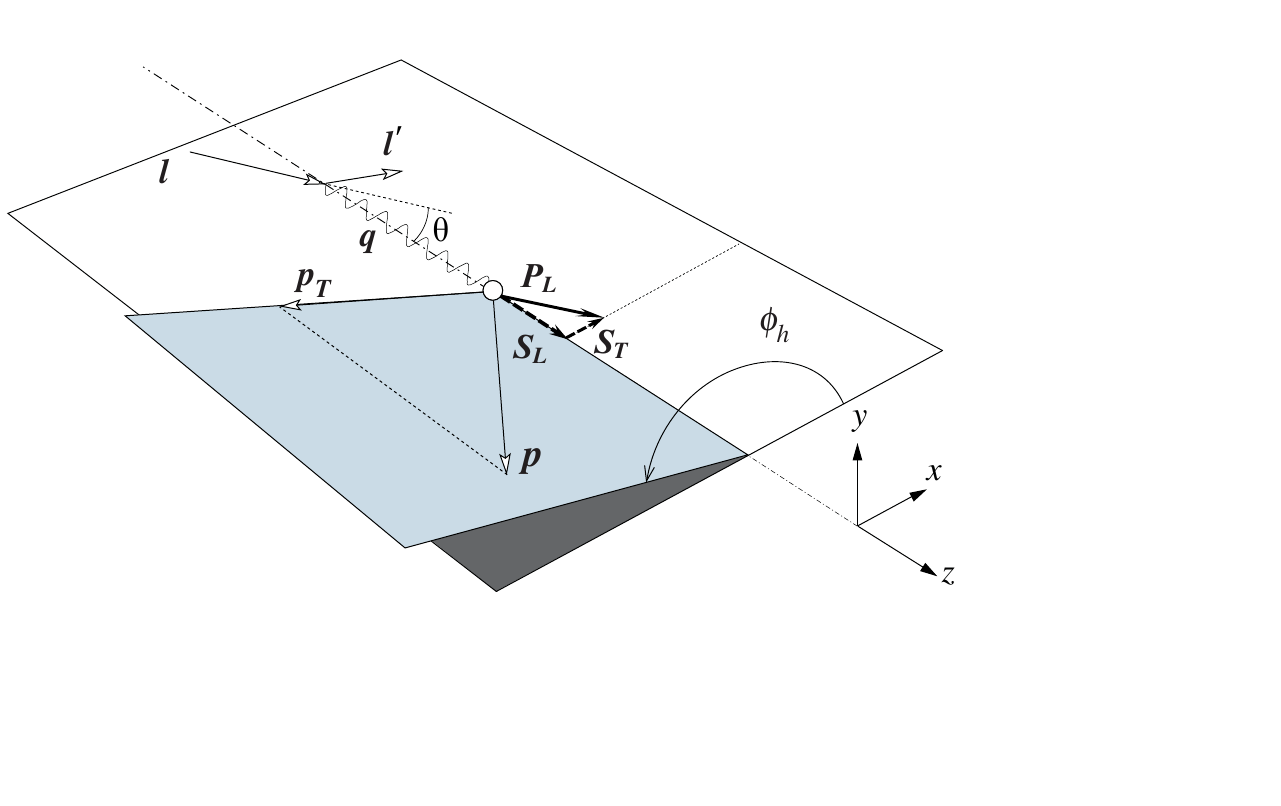}
\caption{The SIDIS process framework and definition of azimuthal angle $\phiH$.}
\label{fig:SIDIS_frame}
\end{wrapfigure}

When applying the TMD factorisation~\cite{Collins:2011zzd} for the SIDIS cross sections, asymmetries are interpreted as convolutions of different TMD PDFs and quark fragmentation functions (FFs) which describe (un)polarized quark fragmentation mechanisms.
In the gamma-nucleon system  ($\gamma^*N$) adopted in this Letter, the general expression for the cross section of unpolarized-hadron production in polarized-lepton SIDIS off a polarized nucleon contains three target spin independent, four longitudinal and eight transverse target-polarization-dependent modulations in the azimuthal angles of the produced hadron and that of the target spin vector~\cite{Kotzinian:1994dv,Bacchetta:2006tn}. Schematic view of the $\gamma^*N$ coordinate system is presented in Fig.~\ref{fig:SIDIS_frame} for the case when nucleon is polarized longitudinally w.r.t. the beam.
It is a right-handed coordinate system defined by the virtual photon direction (z axis), lepton transverse (w.r.t the z-axis) momentum direction (x axis), and respective normal vector (y axis). Symbols $l$, $l'$ and $q$ represent the four-momenta of the incident and outgoing leptons and virtual photon ($\gamma^*$), respectively, while vectors ${\bf p}$ and ${\bf p}_{T}$ denote the momentum and the transverse momentum of the produced hadron $\it h$. The angle $\phiH$ defines the azimuthal orientation of the hadron momentum with respect to the lepton scattering plane calculated about the virtual-photon momentum direction. In the introduced framework, $S_{L}$ ($S_{T}$) is the longitudinal (transverse) component of the target polarization vector ${\bf S}$, while $P_{L}$ stands for the longitudinal component of the target polarization w.r.t. the beam direction.

Choosing the virtual-photon momentum as a reference for definition of the target polarization is a natural basis for theoretical description of $\gamma^*N$ sub-processes. Another reference is the lepton beam direction ($\ell$-axis) which is more relevant from the experimental point of view since the target is usually polarized longitudinally or transversely w.r.t. the beam axis. The values longitudinal and transverse w.r.t. $\ell$-axis polarizations are defined by the experimental setup, whereas $S_{L}$ and $S_{T}$ depend on the orientation of the virtual photon, hence on the kinematics of each individual DIS event. As shown in Fig.~\ref{fig:SIDIS_frame}, the two polarization definitions are related via rotation by the angle $\theta$, the polar angle between the incoming beam direction and the virtual photon direction. As a result, for arbitrary target polarization with respect to the incoming beam direction, the polarization vector defined with respect to the virtual photon direction will have both transverse and longitudinal components (see e.g. Fig.~\ref{fig:SIDIS_frame}). This means that target $\ell$-axis polarization asymmetries (${\ell}$-asymmetries) measured in experiment also contain a mixture of contributions from both $\gamma^*$-axis polarization components (${\gamma^*}$-asymmetries).
In the discussed case when the target is polarized longitudinally w.r.t. incoming lepton, the transverse spin component lies in the lepton-scattering plane leading to $\phiS=0$, or $\phiS=\pi$ (depending on the target polarization direction) and is given by the relation $S_{T}\cos\phiS=-\sin\theta P_L$. Using standard SIDIS notations~\cite{Kotzinian:1994dv,Bacchetta:2006tn} the angle $\theta$ is given by the relation, $\sin\theta = \gamma\,\sqrt{\frac{1-y- \frac{1}{4} y^2 \gamma^2}{1+\gamma^2}}$, where $\varepsilon = (1-y -\frac{1}{4}\slim \gamma^2 y^2)/(1-y +\frac{1}{2}\slim y^2 +\frac{1}{4}\slim \gamma^2y^2)$  is the ratio of longitudinal and transverse photon fluxes and $\gamma = 2 M x/Q$. In COMPASS kinematics $\sin\theta$ is small ($\langle\sin\theta\rangle\approx0.04$, $\sin\theta<0.15$) and thus the $S_{T}$ component is kinematically suppressed.
%
%
%

Following general considerations the SIDIS cross section for $\gamma^*N$-system and longitudinally (w.r.t. lepton beam) polarized target can be written in a following model-independent way~\cite{Kotzinian:1994dv,Bacchetta:2006tn,Diehl:2005pc}:
%
%
%
%
%
\begin{eqnarray}\nonumber
\centering
\label{eq:x_secSIDIS}
&& \hspace*{-1.0cm}\frac{{d\sigma }}{{dxdydzp_{T}dp_{T}d{\phiH}}} \propto \left( {{F_{UU,T}} + \varepsilon {F_{UU,L}}} \right)
%
%
\Bigg\{ 1 + P_{L}-independent\;azimuthal\;asymmetries \\
%
&&\hspace*{2.8cm}+\,{P_{L}}\Big[ \sqrt {2\varepsilon \left( {1 + \varepsilon } \right)}\blue{A_{UL^{\ell}}^{\sin {\phiH}}}\sin {\phiH}
+\,\varepsilon \red{A_{UL^{\ell}}^{\sin {2{\phiH}}}}\sin {2{\phiH}}
%
-\,\green{\underline{A_{UL^{\ell}}^{\sin{3{\phiH}}}\sin {3{\phiH}}}}\Big]\\ \nonumber
&&\hspace*{2.8cm}+\,{P_{L}}\lambda \Big[ \sqrt {1 - {\varepsilon ^2}} \red{A_{LL^{\ell}}}
+\,\sqrt {2\varepsilon \left( {1 - \varepsilon } \right)} \blue{A_{LL^{\ell}}^{\cos {\phiH}}} \cos {\phiH}
%
-\,\green{\underline{A_{LL^{\ell}}^{\cos{2{\phiH}}}\cos{2{\phiH}}}}\Big]\Bigg\}.
\end{eqnarray}
%
Throughout this Letter $A_{XY^{Z}}^{w(\phi_h,\phi_S)}$ asymmetries are defined as the amplitudes of the corresponding azimuthal modulations $w(\phi_h,\phi_S)$ divided by the spin and azimuth-independent part of the SIDIS cross section, the effective proton polarization\footnote{effective polarization includes the dilution factor $f$ which describes the fraction of polarizable material in the target}
and corresponding depolarization factor\footnote{in Eq.~\ref{eq:x_secSIDIS} depolarization factors are the $\varepsilon$-dependent coefficients standing in front of the asymmetries}\footnote{the double-spin asymmetries in addition are corrected for the beam polarization}.
Here, the subscript X indicates the state of the lepton polarization, $\lambda$, (either U-unpolarised, or L-longitudinal), while Y stands for the target polarization (L-longitudinal or T-transverse) and Z indicates the reference axis for the target polarization (${\ell}$ or $\gamma^*$\footnote{for simplicity $Z=\gamma^*$ labeling of $\gamma^*$-LSAs in most of the cases is omitted both in the text and in the plots}).

The cross section in Eq.~\ref{eq:x_secSIDIS} contains six (one azimuth-independent and five azimuthal) target longitudinal-spin-dependent asymmetries (LSAs) which is by two more than in the case of the pure $\gamma^*N$ cross section~\cite{Kotzinian:1994dv,Bacchetta:2006tn}. The two extra terms, $\sin {3{\phiH}}$ and $\cos {2{\phiH}}$ (underlined and marked in green), arise due to non-vanishing transverse $\gamma^*$-axis polarization component. For the same reason other four ${\ell}$-LSAs have an admixture from specific transverse-spin-dependent asymmetries (TSAs). Similarly, ${\ell}$-TSAs measured in SIDIS are getting mixed with LSA-contributions. Introducing such admixtures is a complication for theoretical models and phenomenological global fits which operate with pure $\gamma^*N$-inputs.
However, the experiments performing ${\ell}$-asymmetry measurements with both longitudinal and transverse target polarization can provide direct access to $\gamma^*$-LSAs and TSAs. For instance, using the following transformation~\cite{Diehl:2005pc}:
%
%
\begin{equation}\label{eq:PT2ST_A}
\centering
A_{L^{\ell}}\approx A_{L^{\gamma^*}}-C(\varepsilon,\theta)A_{T^{\ell}}
\end{equation}
%
%
allows to reconstruct $\gamma^*$-LSAs combining the information from measured ${\ell}$-LSAs and ${\ell}$-TSAs.
Following this structure, Table~\ref{tab:PT2ST} lists for each measured ${\ell}$-LSAs from Eq.~\ref{eq:x_secSIDIS} corresponding $\gamma^*$-LSAs, $C(\varepsilon,\theta)$ mixing coefficients and ${\ell}$-TSA admixtures.
\begin{table}[H]
\centering
\renewcommand{\arraystretch}{1.2}
\resizebox{7cm}{!}{%
\begin{tabular}{|c|c|c|c|c|c|c|c|}
\hline
$A_{L^{\ell}}$ &
$A_{L^{\gamma^*}}$ &
$C(\varepsilon,\theta)$ &
$A_{T^{\ell}}$ \bigstrut \\
\hline
\multirow{2}*{\blue{$A_{UL^{\ell}}^{\sin \phi _h }$}} &
\multirow{2}*{\blue{$A_{UL}^{\sin \phi _h }$ }} &
$\sin\theta\frac{1}{ \sqrt {2\varepsilon \left( {1 + \varepsilon } \right)}}$ &
\red{$A_{UT^{\ell}}^{\sin (\phi _h -\phi _s )}$} \bigstrut \\
\cline{3-4}
&&$\sin\theta\frac{\varepsilon}{ \sqrt {2\varepsilon \left( {1 + \varepsilon } \right)}}$ &
\red{$A_{UT^{\ell}}^{\sin (\phi _h +\phi _s -\pi )}$} \bigstrut \\ \hline
\red{$A_{UL^{\ell}}^{\sin 2\phi _h}$} &
\red{$A_{UL}^{\sin 2\phi _h }$} &
$\sin\theta\frac{\sqrt {2\varepsilon \left( {1 + \varepsilon } \right)}}{ \varepsilon}$ &
\blue{$A_{UT^{\ell}}^{\sin (2\phi _h -\phi _s )}$} \bigstrut \\ \hline
\green{\underline{$A_{UL^{\ell}}^{\sin 3\phi _h }$}} &
- &
$\sin\theta$ &
\red{$A_{UT^{\ell}}^{\sin (3\phi _h -\phi _s )}$}  \bigstrut \\ \hline
\blue{$A_{LL^{\ell}}^{\cos \phi _h}$} &
\blue{$A_{LL}^{\cos \phi _h }$} &
$\sin\theta\frac{\sqrt {\left( {1 - {\varepsilon ^2}} \right)}}{ \sqrt {2\varepsilon \left( {1 - \varepsilon } \right)}}$ &
\red{$A_{LT^{\ell}}^{\cos (\phi _h -\phi _s)}$}  \bigstrut \\ \hline
\red{$A_{LL^{\ell}}$} &
\red{$A_{LL}$} &
$\sin\theta\frac{\sqrt {2\varepsilon \left( {1 - \varepsilon } \right)}}{ \sqrt {\left( {1 - {\varepsilon ^2}} \right)}}$ &
\blue{$A_{LT^{\ell}}^{\cos \phi _s }$} \bigstrut \\ \hline
\green{\underline{$A_{LL^{\ell}}^{\cos 2\phi _h }$}} &
- &
$\sin\theta$ &
\blue{$A_{LT^{\ell}}^{\cos (2\phi _h -\phi _s)}$} \bigstrut \\ \hline
\end{tabular}
\renewcommand{\arraystretch}{1}
}
\caption {List of LSAs, their TSA-admixtures and $C(\varepsilon,\theta)$-factors.}
\label{tab:PT2ST}
\end {table}
The $A_{UL}^{\sin(2\phiH)}$ is the only azimuthal LSA arising in the SIDIS cross section at leading order\footnote{in Eq.~\ref{eq:x_secSIDIS} and Table~\ref{tab:PT2ST} \textit{leading-twist} asymmetries are marked in red while the \textit{subleading-twist} ones in blue}. It gives access to the convolution of so far unknown $h_{1L}^{\bot q}$ TMD PDF, which describes transversely polarized quarks in the longitudinally polarized nucleon, and the extensively studied Collins fragmentation function, $H_{1q}^{\bot h}$~\cite{Boglione:2015zyc,Aidala:2012mv}. Following Eq.~\ref{eq:PT2ST_A}, the pure $\gamma^*$-LSA can be extracted by correcting measured asymmetry for the TSA-admixture, which in this case is represented by $A_{UT^{\ell}}^{\sin (2\phi _h -\phi _s )}$ \textit{subleading-twist} asymmetry scaled by $C(\varepsilon,\theta)=\sin\theta\frac{\sqrt {2\varepsilon \left( {1 + \varepsilon } \right)}}{ \varepsilon}$ coefficient (see Table~\ref{tab:PT2ST}).

The next two azimuthal LSAs, $A_{UL}^{\sin(\phiH)}$ and $A_{LL}^{\cos(\phiH)}$, enter the cross-section at subleading level (additional dynamic $Q^{-1}$ suppression). Both asymmetries are mixed with LO TSAs. The $A_{UL}^{\sin(\phiH)}$ is mixed with $A_{UT^{\ell}}^{\sin (\phi _h -\phi _s )}$ (Sivers) and $A_{UT^{\ell}}^{\sin (\phi _h +\phi _s -\pi )}$ (Collins) TSAs, while $A_{LL}^{\cos(\phiH)}$ LSA is getting admixture from $A_{LT^{\ell}}^{\cos (\phi _h -\phi _s)}$ asymmetry.

At the \textit{twist-3} level polarized structure functions associated to these two LSAs include four individual contributions given by couplings of specific \textit{twist-2} and \textit{twist-3} PDFs and FFs~\cite{Bacchetta:2006tn}. Keeping only the contributions involving \textit{twist-2} FFs, the $A_{UL}^{\sin(\phiH)}$ asymmetry is described by a combination of two such couplings: $xh_{L}^{q} \otimes H_{1q}^{\bot h}$ and $xf_{L}^{\bot q} \otimes D_{1q}^h$, where $D_{1q}^h$ is the ordinary fragmentation function.
The $xf_{L}^{\bot q}$ is a T-odd TMD PDF that can be interpreted as the \textit{twist-3} analog of the Sivers function, while $xh_{L}^{q}$ is a chiral-odd PDF which enters in convolution with Collins FF.
The contributions coming from both \textit{twist-3} PDFs have been recently calculated in two different spectator-diquark models~\cite{Lu:2014fva}. Applying the widely adopted simplification known as Wandzura-Wilczek approximation (WWA)~\cite{Wandzura:1977qf} the $A_{UL}^{\sin(\phiH)}$ simplifies to just one \textit{twist-2} constituent, $h_{1L}^{\bot q}\otimes H_{1q}^{\bot h}$. The aforementioned $A_{UL}^{\sin(2\phiH)}$ asymmetry is related to the same convolution and thus the measurement of these two LSAs can serve as an important contribution for the study of relevant \textit{twist-2} and \textit{twist-3} PDFs and the general validity of WWA.

As for the $A_{LL}^{\cos(\phiH)}$ asymmetry, it is related to the $xe_{L}^{q} \otimes H_{1q}^{\bot h}$ and $xg_{L}^{\bot q} \otimes D_{1q}^h$ couplings which within WWA simplify to $g_{1L}^{q} \otimes D_{1q}^h$, where $g_{1L}^{q}$ is the TMD helicity PDF. This allows to link this LSA to the azimuth-independent $A_{LL}$ LO double-spin asymmetry related to the same $g_{1L}^{q} \otimes D_{1q}^h$ convolution. At COMPASS, in particular, we study the hadron transverse momentum $p_{T}$ dependence of $A_{LL}$ LSA. As it was suggested in~\cite{Anselmino:2006yc}, behaviour of the $A_{LL}$ asymmetry as a function of $p_{T}$ can be used to disentangle different scenarios of transverse momentum dependence of the $g_{1L}^{q}$ TMD PDF.

The last two LSAs, $A_{UL^{\ell}}^{\sin 3\phi _h }$ and $A_{LL^{\ell}}^{\cos 2\phi _h }$, are induced due to the non-vanishing target transverse $\gamma^*$-axis polarization component and are identical to $\sin\theta$-scaled $A_{UT^{\ell}}^{\sin (3\phi _h -\phi _s )}$ and $A_{LT^{\ell}}^{\cos (2\phi _h -\phi _s)}$ TSAs.

All aforementioned LSAs have been studied by HERMES and CLAS collaborations with proton target~\cite{Airapetian:2005jc,Avakian:2010ae} and by COMPASS experiment on deuteron target~\cite{Adolph:2016vou}. Similar measurements were done also for the TSAs~\cite{Alekseev:2008aa,Airapetian:2009ae,Adolph:2012sp,Adolph:2012sn,Parsamyan:2014uda}.
In this Letter preliminary COMPASS results for all six proton longitudinal-spin-dependent asymmetries (four ${\gamma^*}$-LSAs and two ${\ell}$-LSAs) will be presented. The asymmetries are corrected for TSA-contributions and carefully studied for possible systematic effects. Obtained results are given in various kinematic representations in order to provide more detailed input for relevant studies of involved TMD PDFs and FFs.

\section{Data analysis}
\label{sec:data_analysis}
The present study is based on COMPASS data taken during nine weeks of data-taking in 2007 and eleven weeks in 2011 using a naturally polarized $\mu^+$-beam of nominal energy of 160 \gvc and 200 \gvc, respectively, and a three celled (30 cm, 60 cm, 30 cm) longitudinally polarized ammonia target.
The data of both years were scrutinized separately by various stability and quality tests.

In order to ensure the DIS-regime the negative square of the virtual photon's four momentum, $Q^{2}$, is required to fulfill the condition $Q^{2} > 1\,(GeV/c)^2$. Contributions from exclusive nucleon resonance production are suppressed by the restriction on the invariant mass of the $\gamma^*N$ system, $W > 5\,GeV/c^{2}$.
The Bjorken scaling variable is limited to $ 0.0025 < x < 0.7$.
Furthermore, the fractional energy of the virtual photon is limited to $ 0.1 < y < 0.9$. Here, the upper $y$ cut rejects events, where radiative corrections become not negligible, while the lower $y$ cut removes the elastic tail and rejects events, where halo or background muons are misidentified as scattered muons.

To ensure a good resolution of the measured angle $\phi_{h}$, the transverse component of the momentum of the hadron is required to satisfy the criteria $p_{T} > 0.1\,{GeV/c}$.
The fractional energy of each hadron is limited to $0.1 < z < 1.0$. The asymmetries were extracted for three $z$ intervals: $z>0.1$, $z>0.2$ and $0.1<z<0.2$.

\begin{figure*}
\centering
\includegraphics[width=1.0\textwidth]{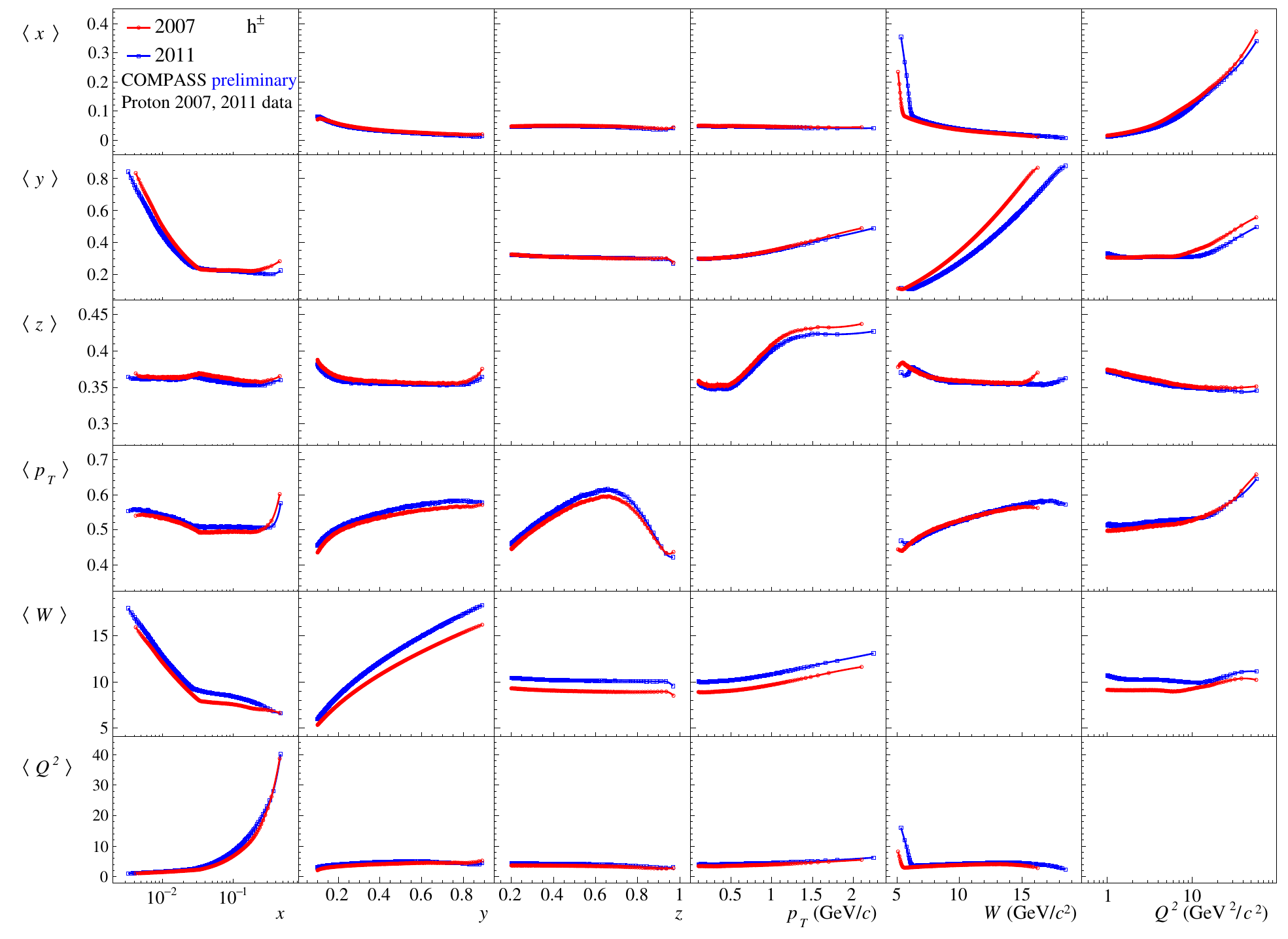}
\caption{Mean values of kinematic variables in different kinematic bins for 2007 and 2011.}
\label{fig:Kmap}
\end{figure*}

After all cuts there are in total about \SI{~e8} charged hadrons in range $z>0.1$ for 2007 and \SI{~0.8e8} for 2011.
The multidimensional dependences of mean values of important kinematic variables for charged hadrons for both years are shown in Fig.~\ref{fig:Kmap}. The different beam energies in 2007 and 2011 explain some relatively small deviations in $x$, $Q^2$, $y$, and $W$. In particular, it is evident that higher beam energy in 2011 allows to access lower values of $x$ and larger values of $W$.
In general, variables, considered for the extraction of asymmetries, show comparable values, which allowed to merge the two samples and to study the asymmetries for combined 2007$-$2011 data set.

\begin{figure*}
\centering
\includegraphics[width=1.0\textwidth]{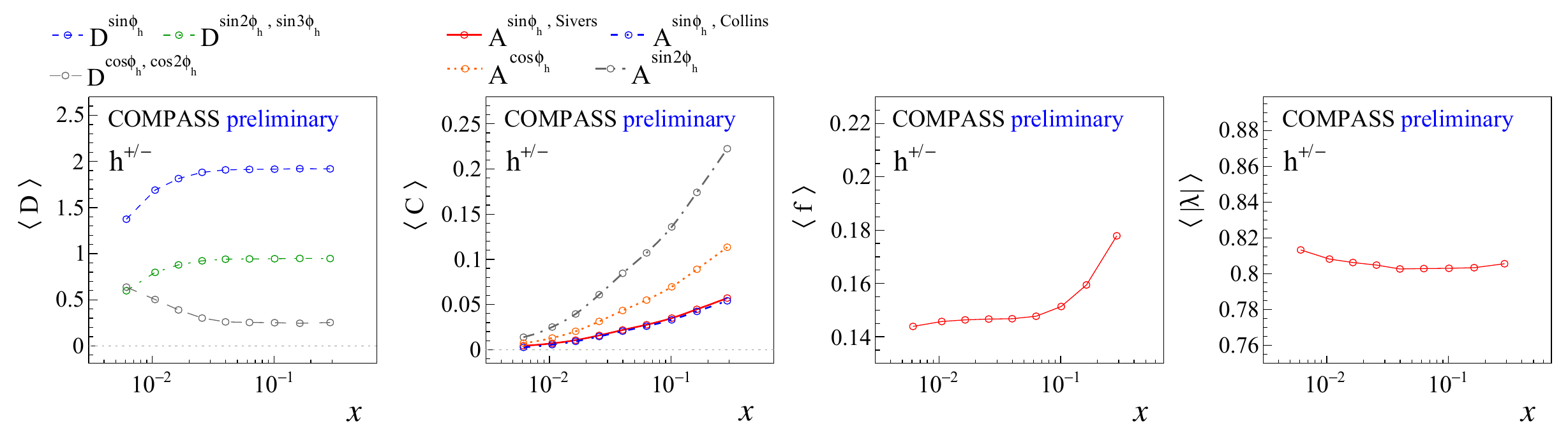}
\caption{From left to right: average $D$, $C$, $f$ and $|\lambda|$ factors as a function of $x$.}
\label{fig:fPDyC}
\end{figure*}

All six ${\ell}$-LSAs entering in Eq.~\ref{eq:x_secSIDIS} are extracted in bins of $x$, $z$ and $p_{T}$ using one-dimensional "Quadruple Ratio" method~\cite{Alekseev:2008aa}. Obtained asymmetries are then corrected for depolarization factors, dilution factor and target and beam (only double spin LSAs) polarizations.
The correction factors are shown in Fig.~\ref{fig:fPDyC} as extracted from the data.

Following Eq.~\ref{eq:PT2ST_A} pure $\gamma^*$-LSAs, $A_{LL}$, $A_{UL}^{\sin(\phiH)}$, $A_{UL}^{\sin(2\phiH)}$ and $A_{UL}^{\cos(\phiH)}$ have been calculated by correcting corresponding measured ${\ell}$-LSAs for the appropriate TSA-admixtures
\begin{wrapfigure}{r}{8.0cm}
\centering
\includegraphics[width=0.53\textwidth]{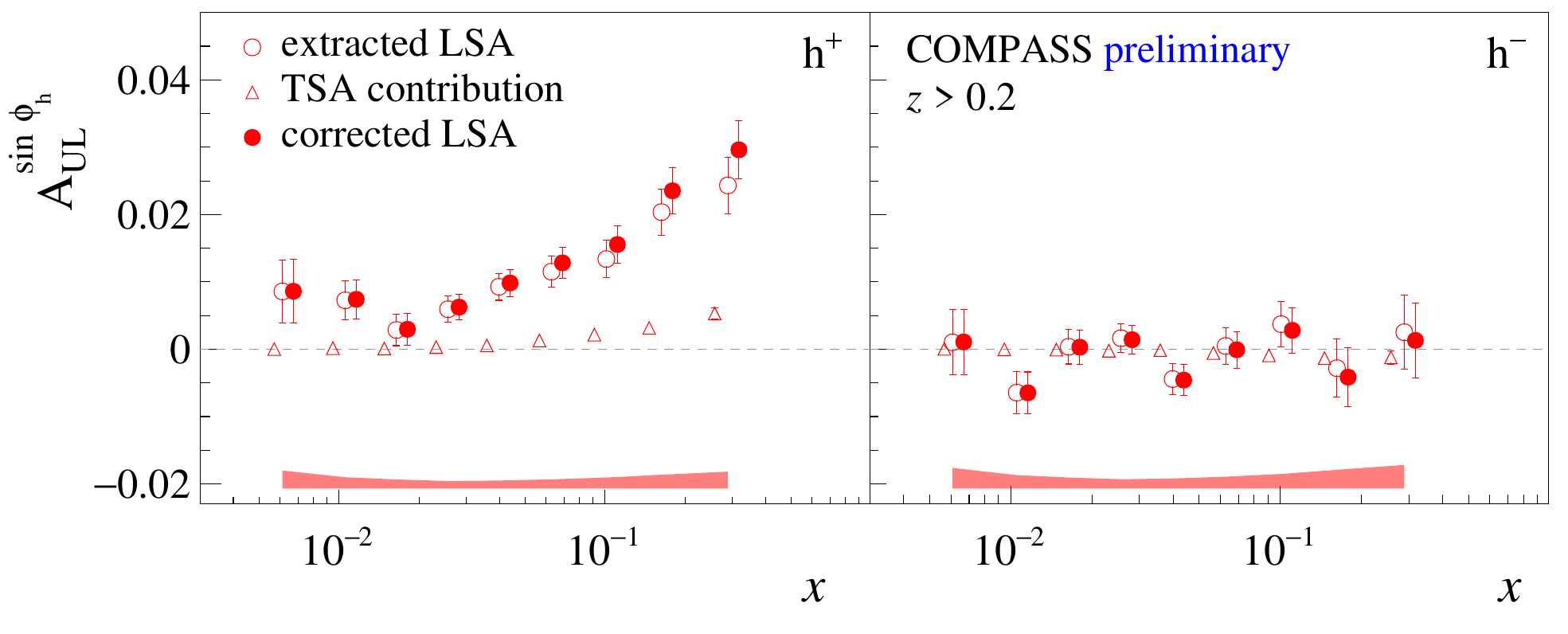}
\caption{$A_{UL}^{\sin(\phiH)}$ - transition from ${\ell}$- to $\gamma^*$-asymmetry.}
\label{fig:A_UL1mix}
\end{wrapfigure}
specified in Table~\ref{tab:PT2ST}.
The impact of such transition is shown in Fig.~\ref{fig:A_UL1mix} on the example of the $A_{UL}^{\sin(\phiH)}$ LSA.
Measured and corrected asymmetries are shown together with the TSA-component which in this case is given by Sivers and Collins TSAs~\cite{Adolph:2012sp,Adolph:2012sn}.

A set of systematic studies has been performed to test the stability of obtained results over the time and to identify possible systematic biases via study of various false asymmetries.
The results of false-asymmetry tests coupled with the outcome of periods compatibility test and results of the monitoring of the mean values of the asymmetries on period-by-period basis, have not given any critical indications for any of the periods of the 2007 and 2011 data taking.
Final results were built combining the asymmetries obtained from 2007 and 2011 taking into account corresponding statistical and systematic uncertainties.

\section{Results and Discussion}
\label{sec:results}

\begin{figure}[]
\centering
\includegraphics[width=0.495\textwidth]{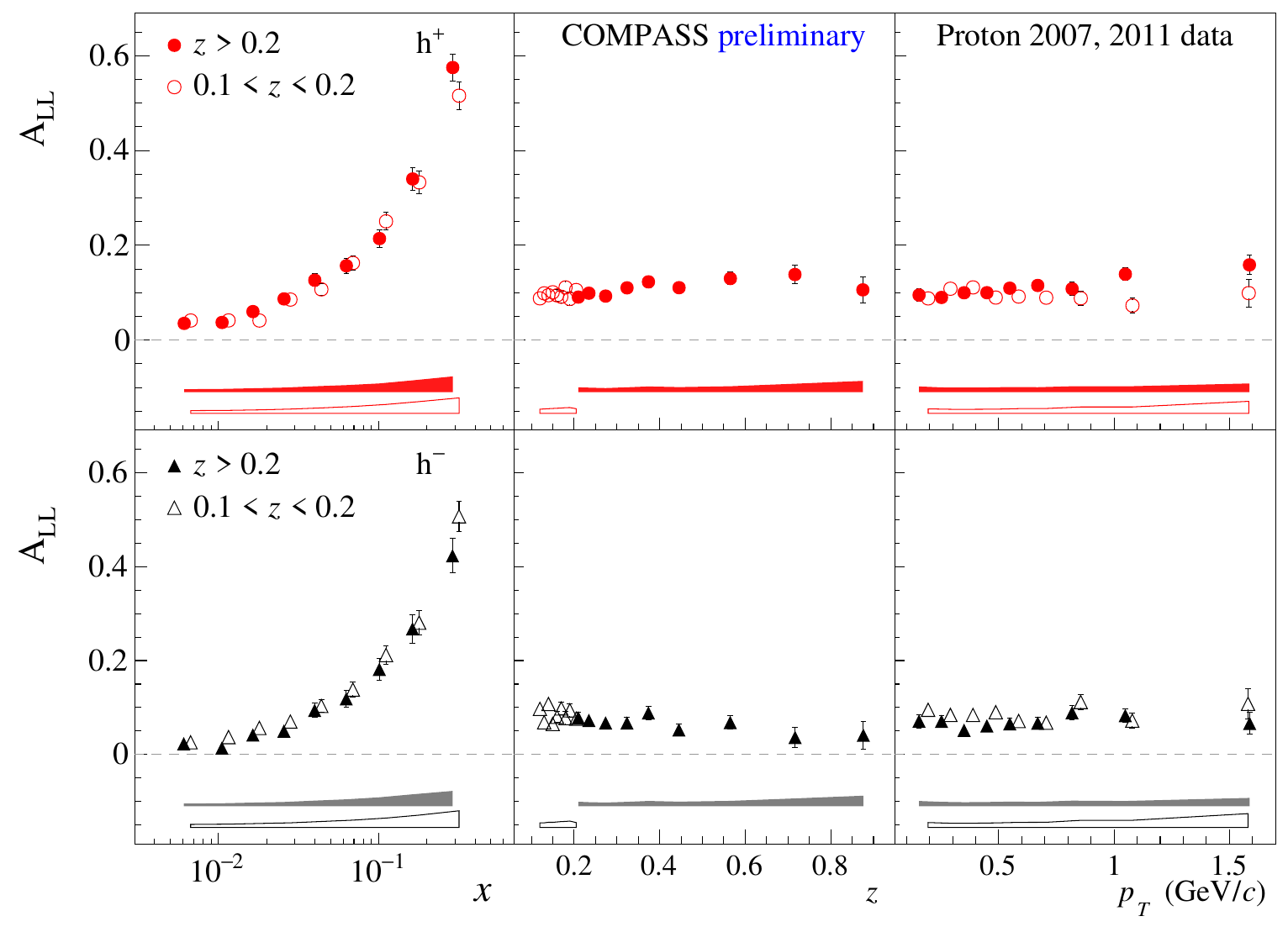}
\includegraphics[width=0.495\textwidth]{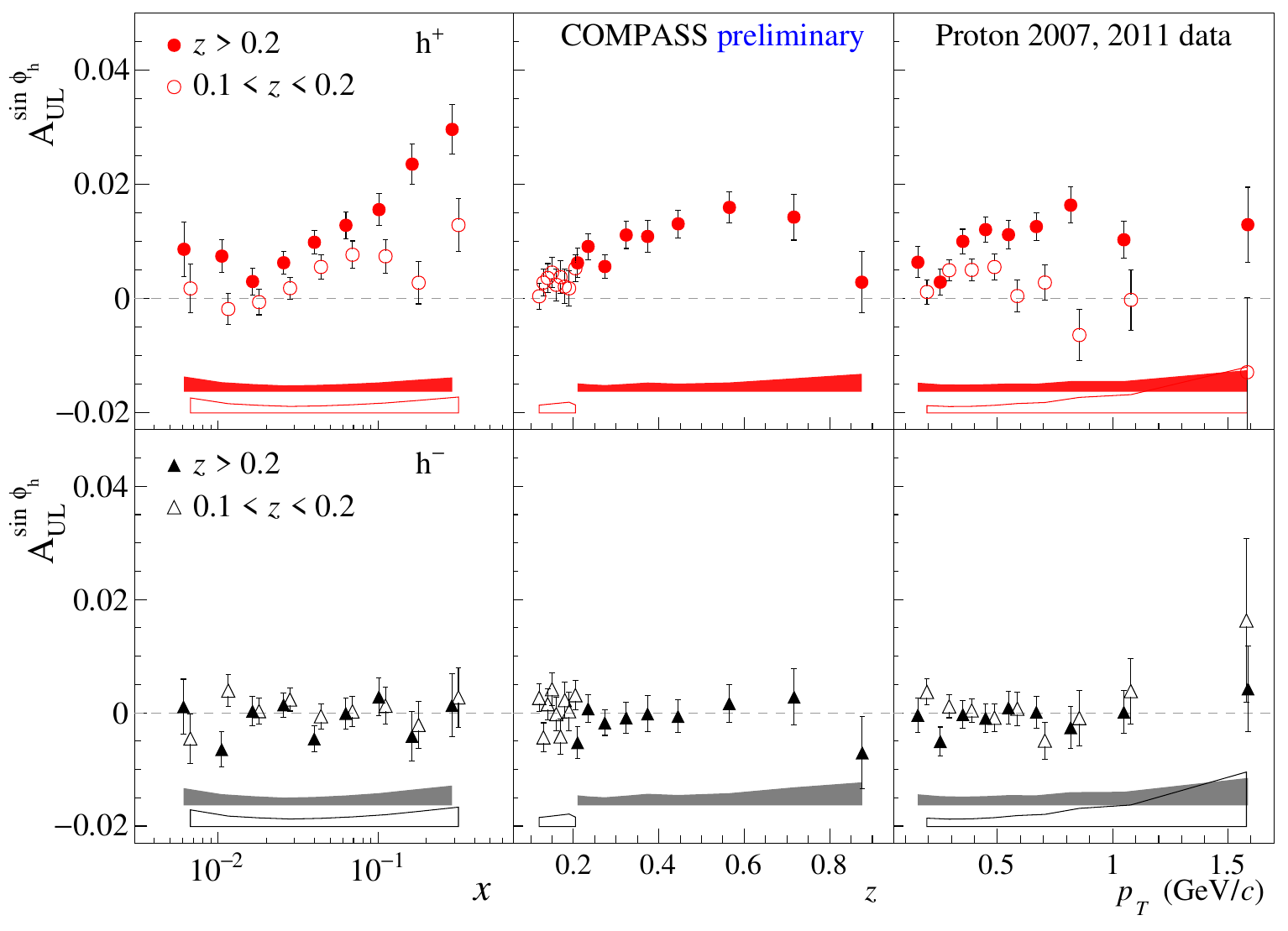}
\includegraphics[width=0.495\textwidth]{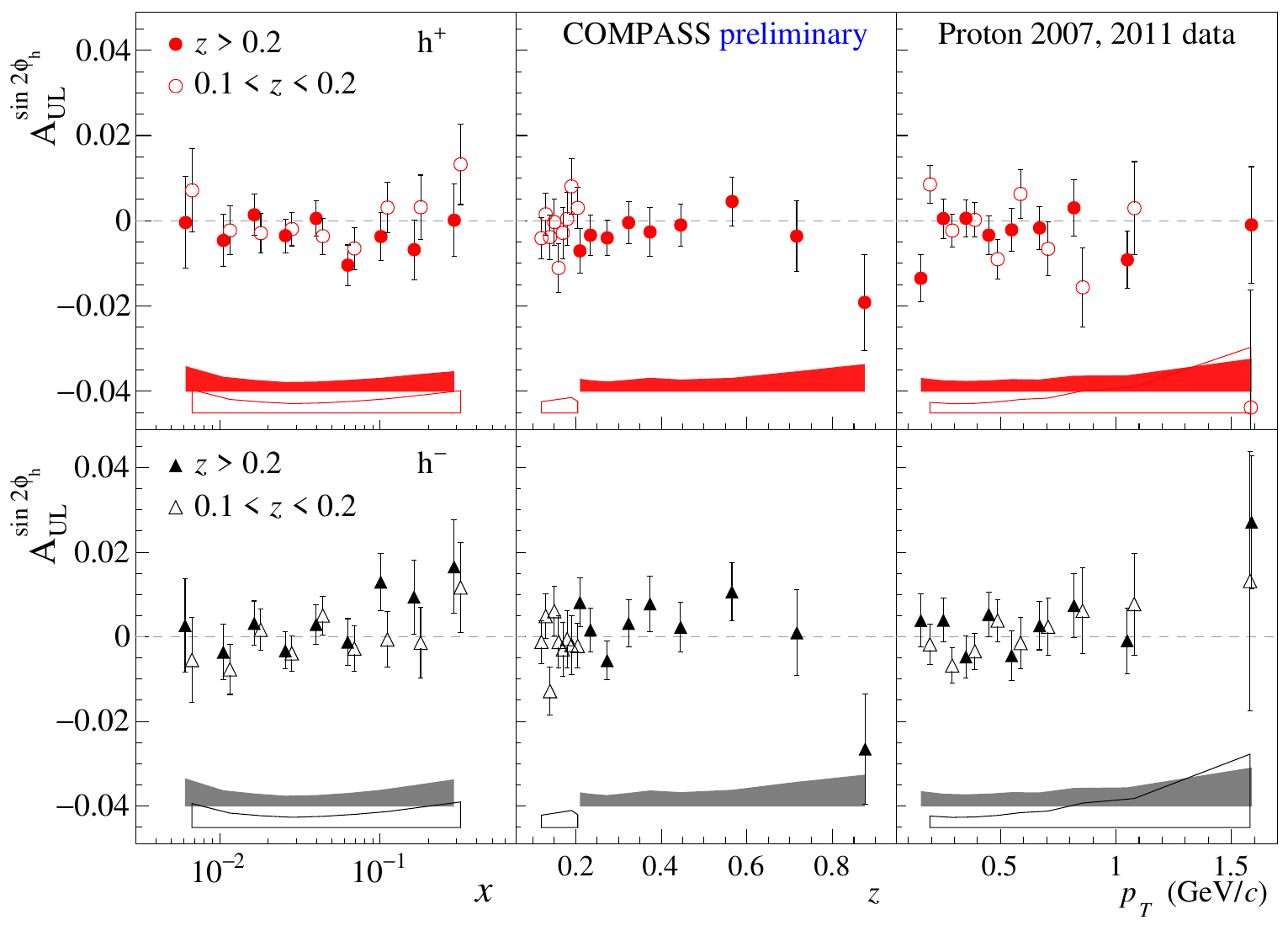}
\includegraphics[width=0.495\textwidth]{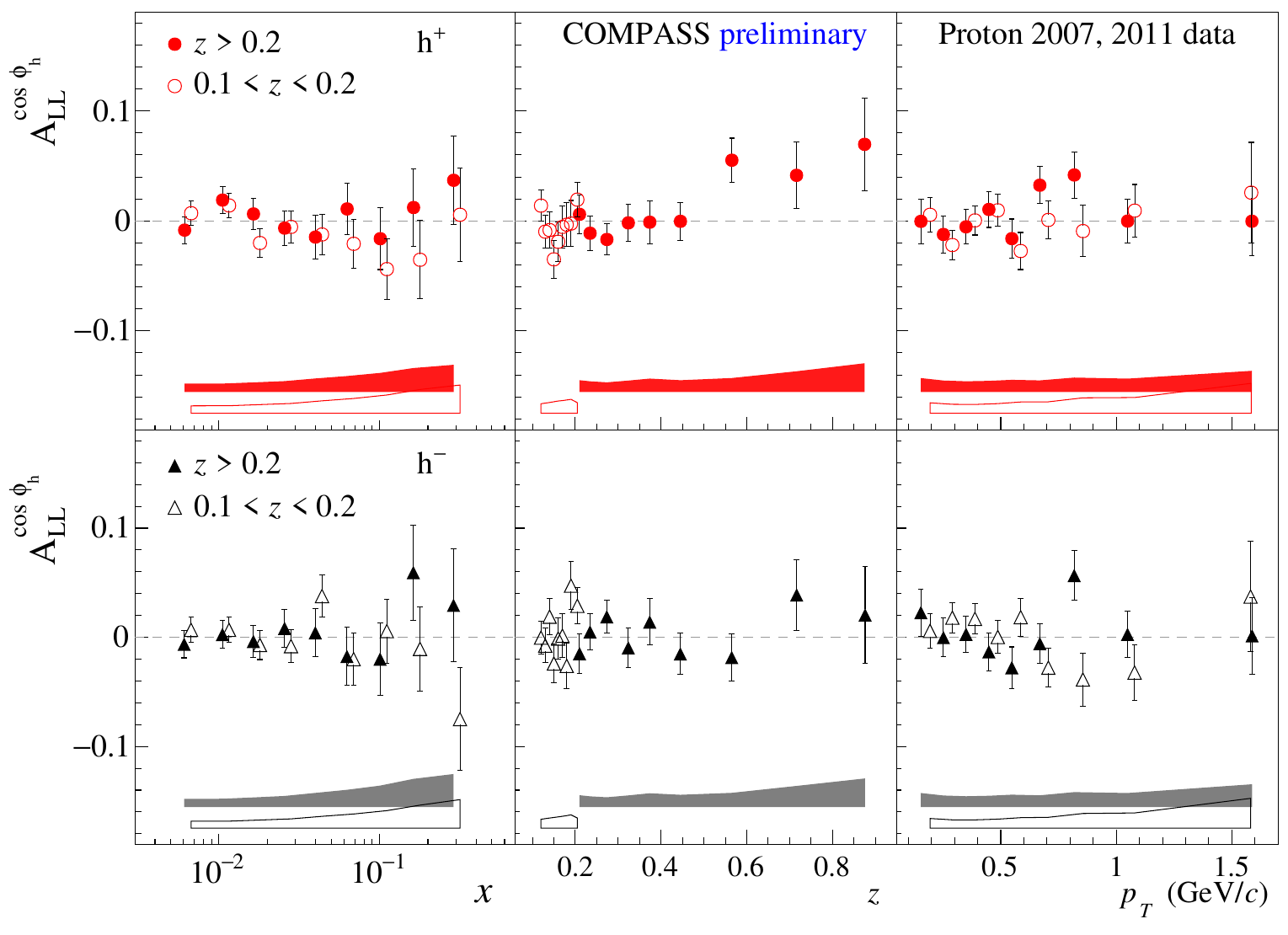}
\caption{Kinematic dependences of $A_{LL}$, $A_{UL}^{\sin(\phiH)}$, $A_{UL}^{\sin(2\phiH)}$ and $A_{UL}^{\cos(\phiH)}$ LSAs for $z>0.2$ and $0.1<z<0.2$ ranges.}
\label{fig:LSA_1h}
\end{figure}

\begin{figure}[]
\centering
\includegraphics[width=0.495\textwidth]{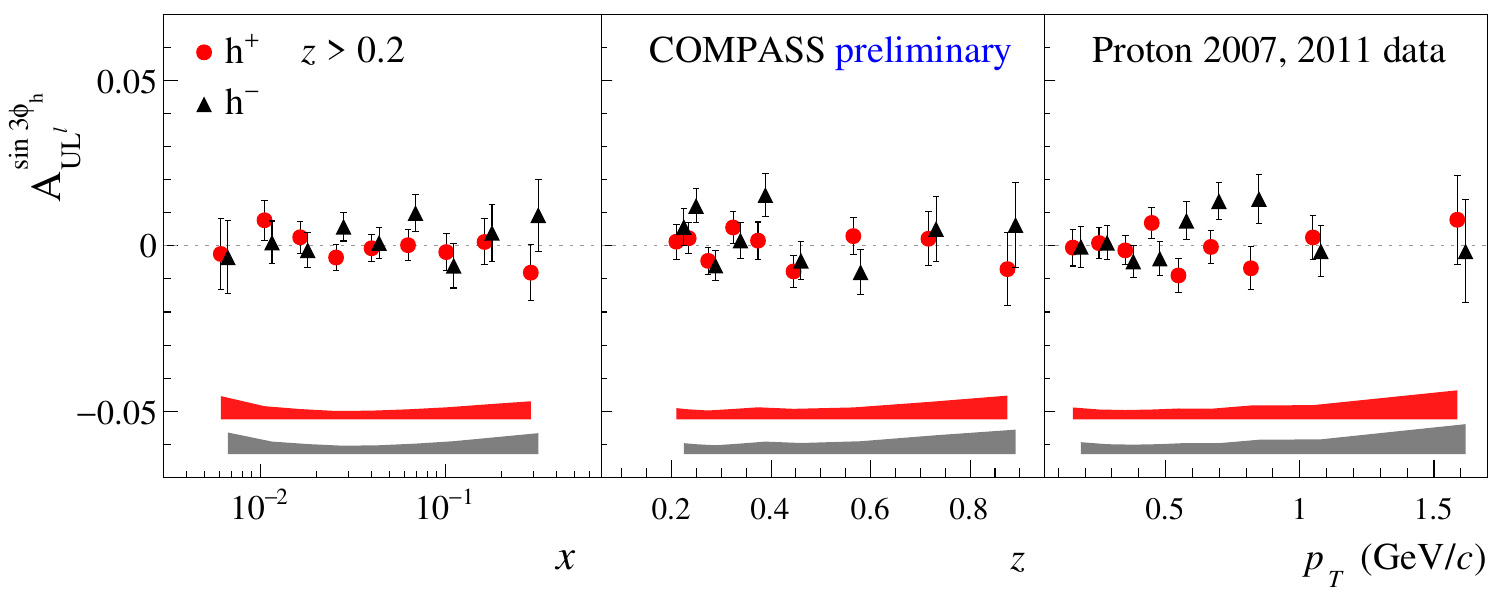}
\includegraphics[width=0.495\textwidth]{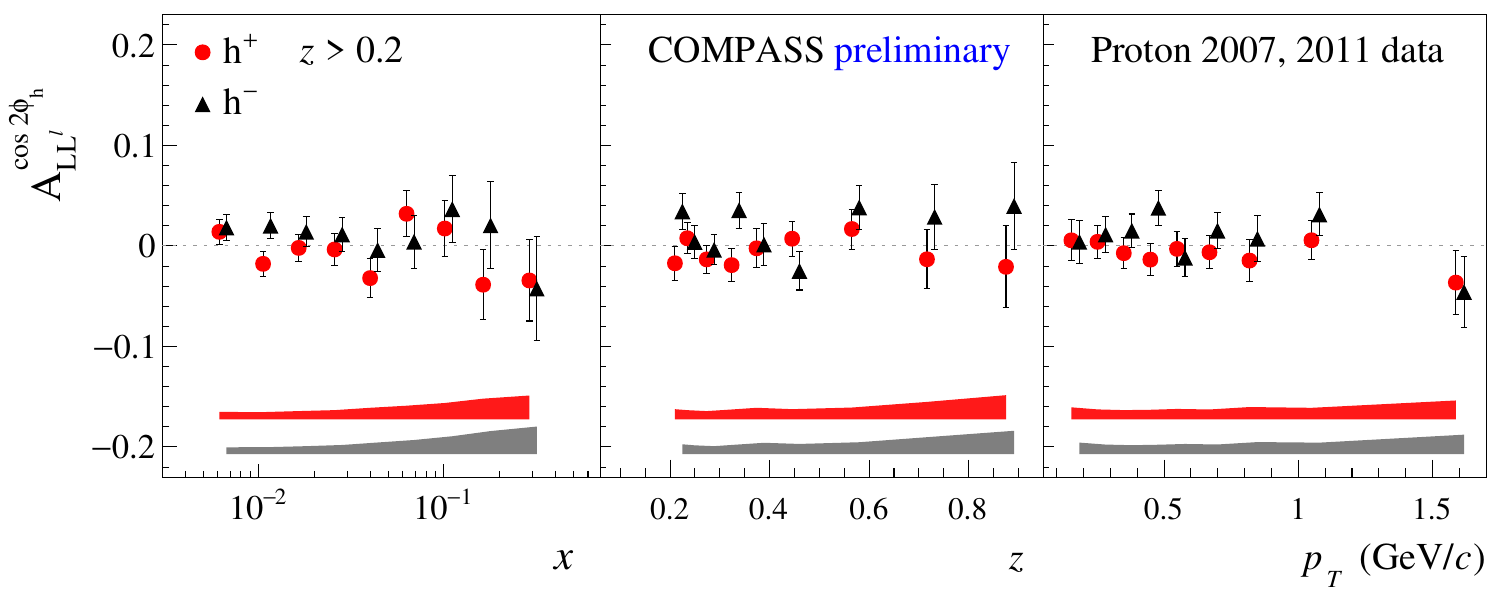}
\caption{Kinematic dependences of $A_{UL^{\ell}}^{\sin(3\phiH)}$ and $A_{UL^{\ell}}^{\cos(2\phiH)}$ LSAs.}
\label{fig:lN_LSA_1h}
\end{figure}

\begin{figure}[]
\centering
\includegraphics[width=0.3\textwidth]{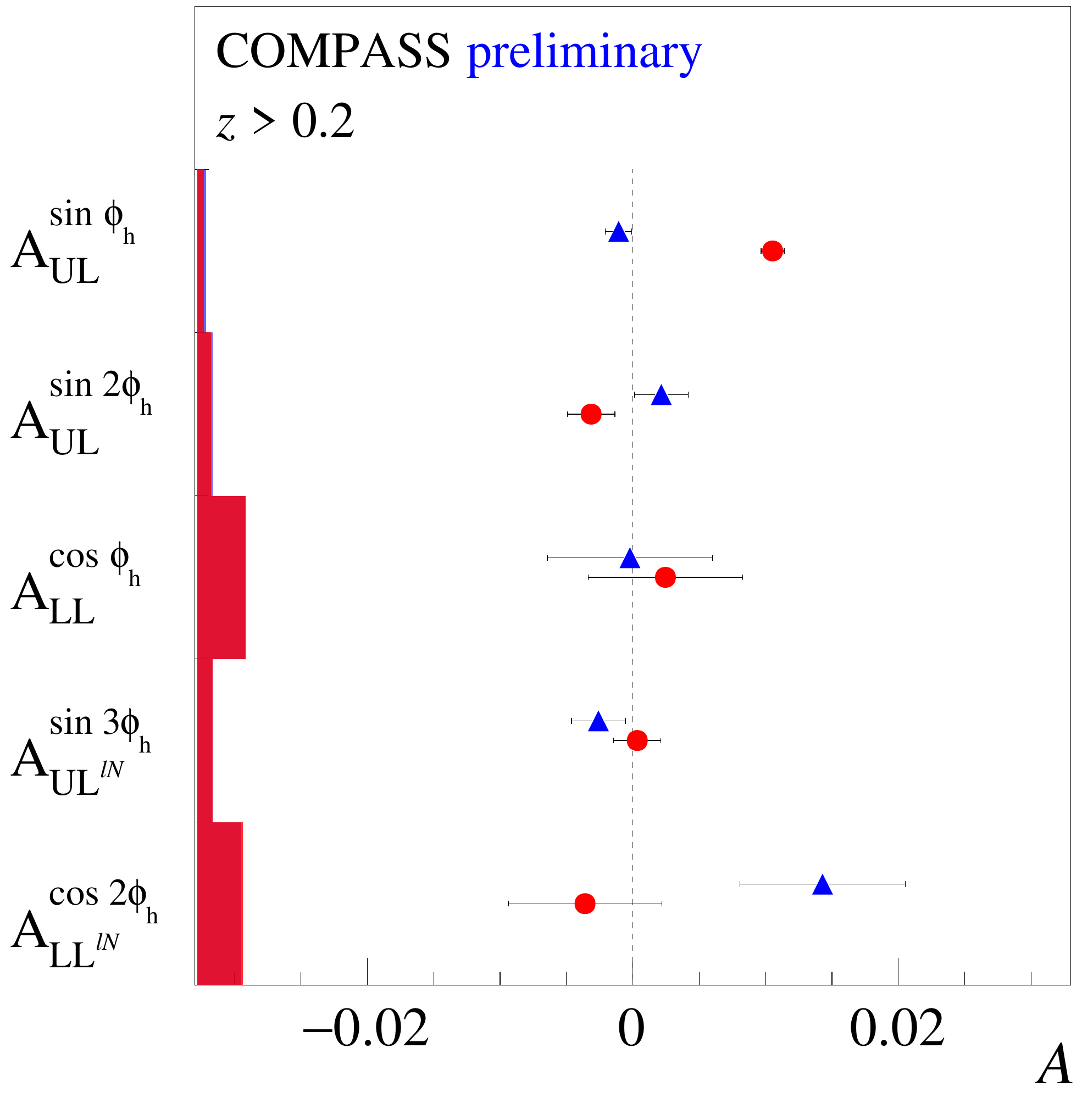}
\hspace{1cm}
\includegraphics[width=0.3\textwidth]{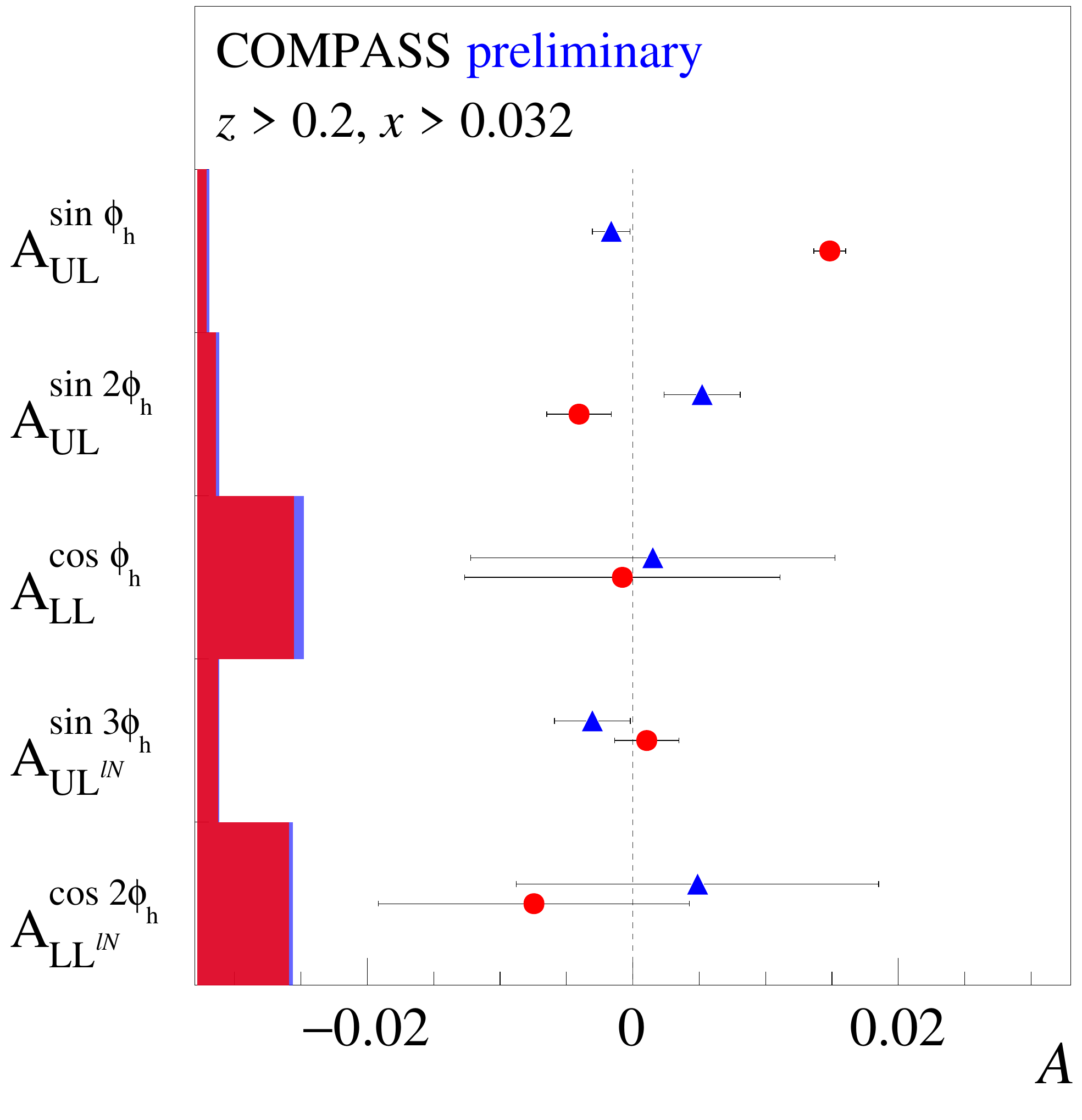}
\caption{The azimuthal LSAs extracted from COMPASS SIDIS data of 2007 and 2011 are shown for the overall $x$ range (top) and for $x>0.032$ (bottom), after averaging over all other kinematic dependences.}
\label{fig:LSA_1h_Mean}
\end{figure}

\begin{figure}[]
\centering
\includegraphics[width=0.495\textwidth]{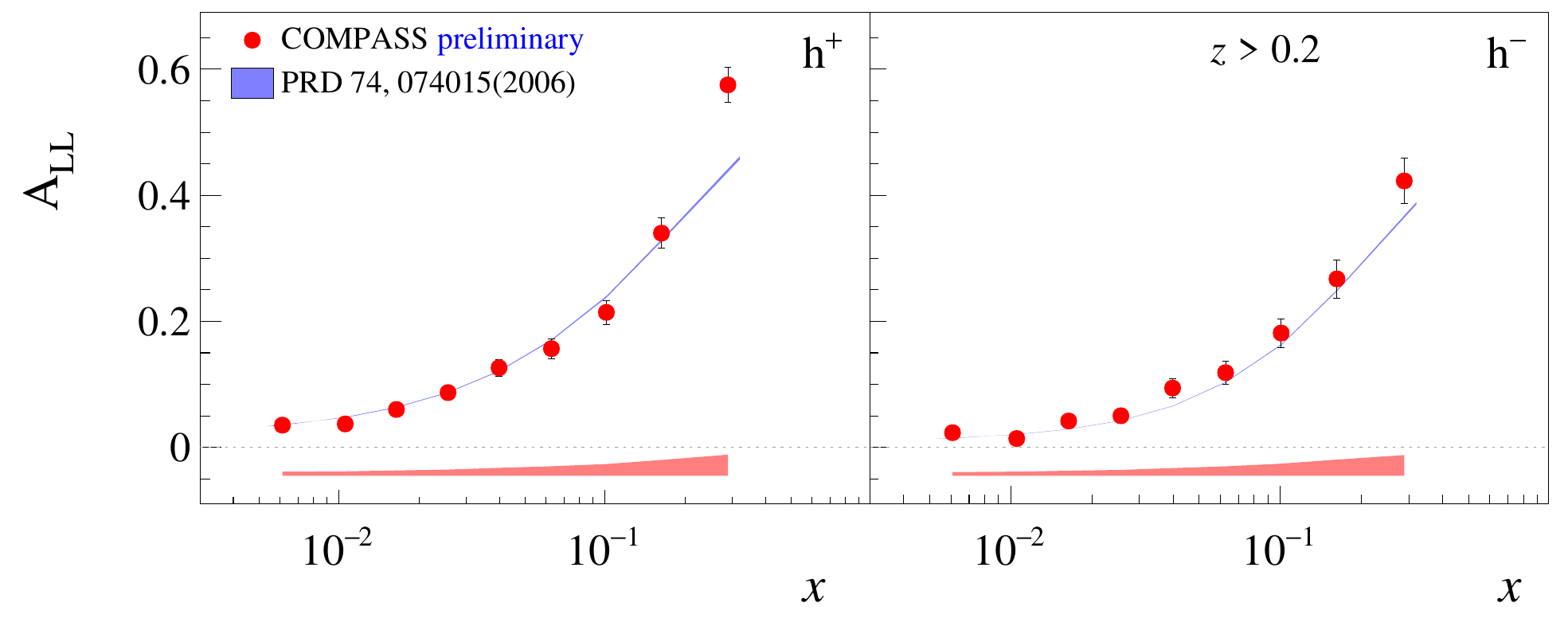}
\includegraphics[width=0.495\textwidth]{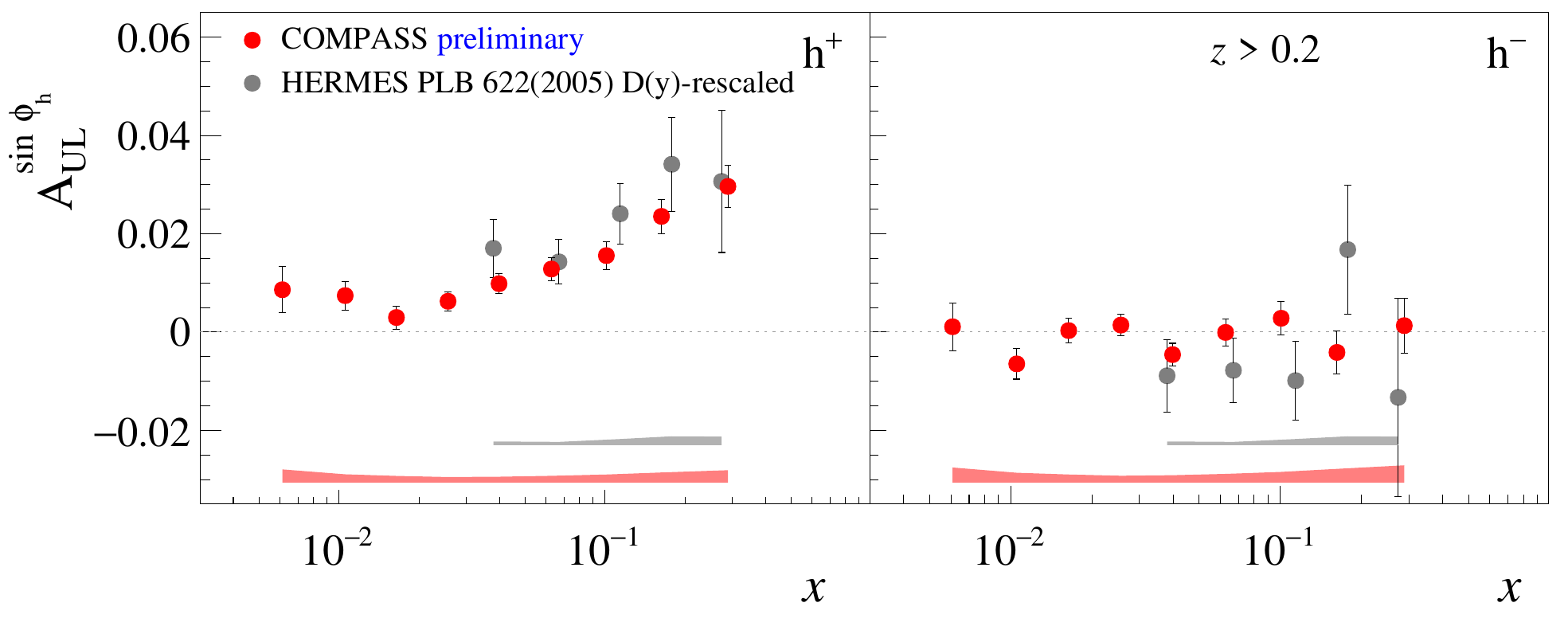}
\includegraphics[width=0.495\textwidth]{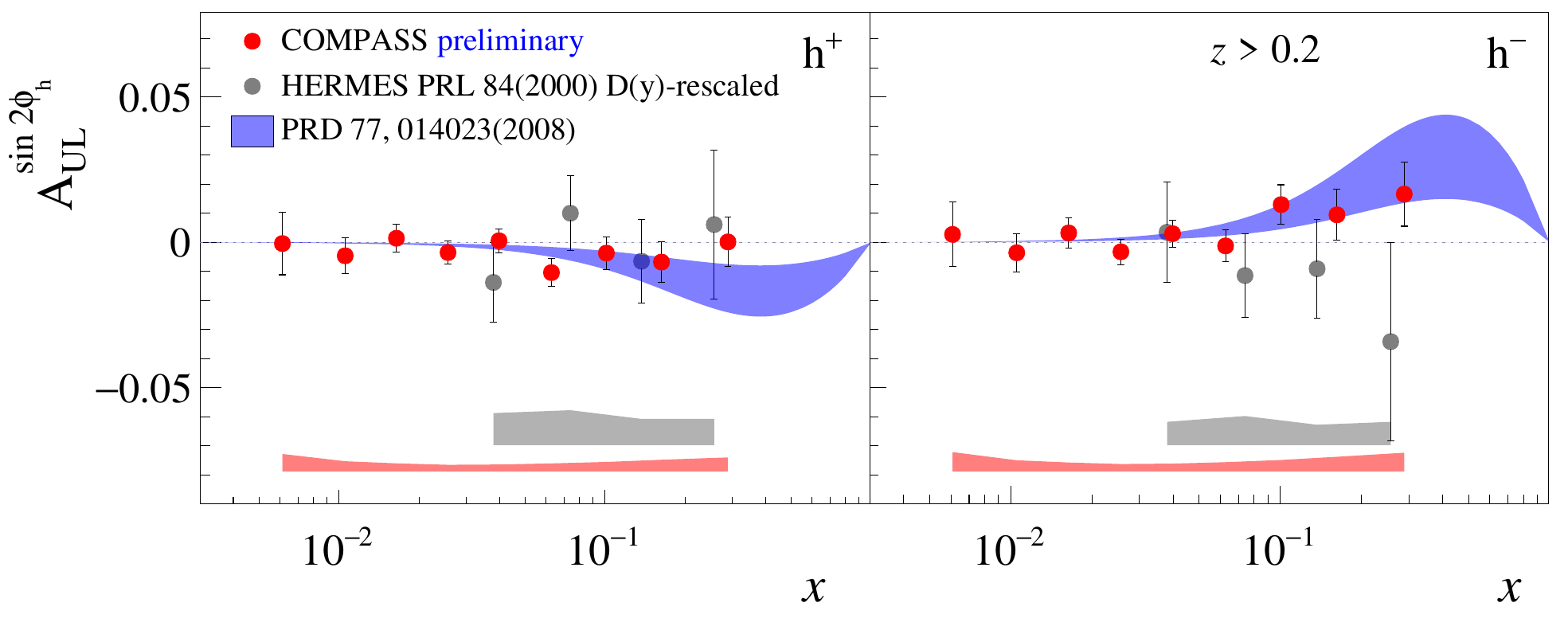}
\includegraphics[width=0.495\textwidth]{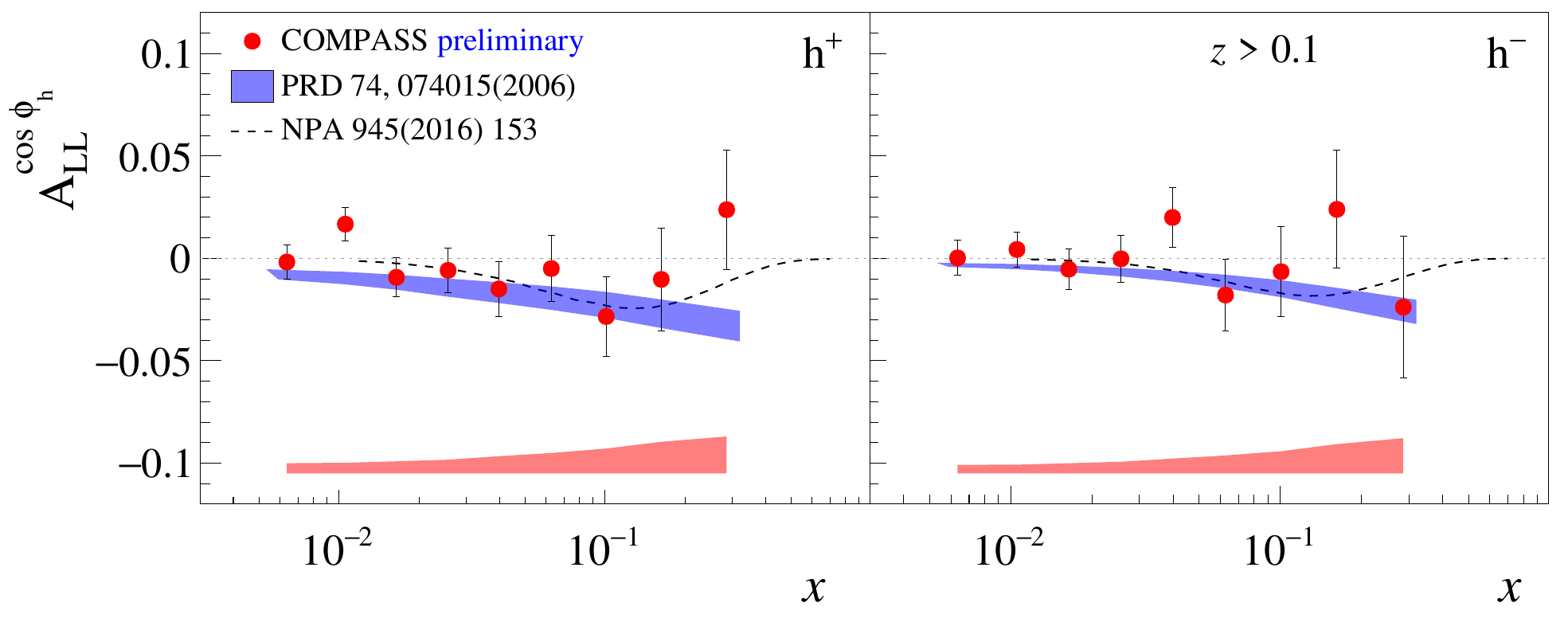}
\caption{The x-dependence of $A_{LL}$, $A_{UL}^{\sin(\phiH)}$, $A_{UL}^{\sin(2\phiH)}$ and $A_{UL}^{\cos(\phiH)}$ LSAs. The results from HERMES (re-scaled by the corresponding $D(y)$ factors) as well as available model  predictions~\cite{Anselmino:2006yc,Avakian:2007mv,Mao:2016hdi} are shown for a comparison.}
\label{fig:LSA_1h_X_th}
\end{figure}

The six LSAs that are extracted from COMPASS SIDIS data of 2007 and 2011 in this analysis are shown in Figs.~\ref{fig:LSA_1h}--\ref{fig:LSA_1h_X_th}. The statistical uncertainties are represented by the error bars and the systematic uncertainties are indicated by color bands on the bottom (or on the left side, for Fig.~\ref{fig:LSA_1h_Mean}) of each plot.
The LSAs, $A_{LL}$, $A_{UL}^{\sin(\phiH)}$, $A_{UL}^{\sin(2\phiH)}$ and $A_{UL}^{\cos(\phiH)}$ extracted for two different $z$-ranges ($0.1<z<0.2$ and $z>0.2$) are shown in Fig.~\ref{fig:LSA_1h} as a function of the variables $x$, $z$ and $p_{T}$. The kinematic dependences of the $A_{UL^{\ell}}^{\sin(3\phiH)}$ and $A_{UL^{\ell}}^{\cos(2\phiH)}$ LSAs are shown separately in the Fig.~\ref{fig:lN_LSA_1h}.
Results for all five azimuthal LSAs integrated over the entire kinematic range are shown in Fig.~\ref{fig:LSA_1h_Mean}. The two panels correspond to different $x$-ranges (top panel: full $x$-range, bottom panel: $x>0.032$). Finally, in Fig.~\ref{fig:LSA_1h_X_th} the LSAs obtained by COMPASS ($x$-dependence only) are superimposed with available theoretical predictions~\cite{Anselmino:2006yc,Avakian:2007mv,Mao:2016hdi} and the experimental results obtained previously by the HERMES collaboration~\cite{Airapetian:2005jc}.

The azimuth-independent $A_{LL}$ asymmetry shown in Fig.~\ref{fig:LSA_1h} (top left panel) as expected exhibits strong $x$-dependence well described by models, see Fig.~\ref{fig:LSA_1h_X_th}. At the same time the $z$ and $p_{T}$ dependences of the LSA appear to be rather flat. As suggested in Ref.~\cite{Anselmino:2006yc}, flat behaviour of the $A_{LL}$ as a function of $p_{T}$ could point to the similarity of the average quark transverse motion inside unpolarized and longitudinally polarized nucleons.

Apart from the well-known large effect for $A_{LL}$-asymmetry, all other LSAs according to various model predictions are predicted to be small both at COMPASS and HERMES kinematics, reaching maximal values of 4-5 percent in the region of large $x$ ($x>0.1$)~\cite{Anselmino:2006yc,Avakian:2007mv,Mao:2016hdi}.
At COMPASS the only azimuthal LSA which exhibits clear non-zero behaviour is the $A_{UL}^{\sin(\phi_h)}$ subleading amplitude. Presented in Fig.~\ref{fig:LSA_1h_Mean} mean $A_{UL}^{\sin(\phi_h)}$ LSA for positive hadrons is about nine standard deviations above zero. This measurement for positive hadrons confirms with much higher precision the previous non-zero observations made by the HERMES collaboration~\cite{Airapetian:2005jc} (see Fig.~\ref{fig:LSA_1h_X_th} top right panel). It worth to note that the effect is expected to be larger at HERMES due to smaller values of $Q$ in the given $x$-bin (because of its subleading nature, the $A_{UL}^{\sin(\phi_h)}$ LSA is suppressed by $Q^{-1}$ pre-factor). In addition, as one can conclude from Fig.~\ref{fig:LSA_1h}, the asymmetry strongly depends on choice of the $z$-range. Therefore, the comparison with HERMES results is not straightforward. Finally, obtained results for $A_{UL}^{\sin(\phi_h)}$ appear to be in qualitative agreement with theoretical predictions from Refs.~\cite{Lu:2014fva} done for COMPASS kinematics.

The $A_{UL}^{\sin(2\phiH)}$ term is the only azimuthal LSA arising in the SIDIS cross section at leading order and thus doesn't suffer from $Q^{-1}$ suppression-factor. Nevertheless, the asymmetry comes with kinematic pre-factor which effectively leads to relative $\sim|\vec{p_{T}}|$-suppression of the effect.
At large $x$ there are some hints for a small and statistically unclear (about 1.5-2 standard deviations away from zero, see Fig.~\ref{fig:LSA_1h_Mean}) "mirror-symmetric"  effect for oppositely charged hadrons. This agrees with model calculations~\cite{Avakian:2007mv} predicting small asymmetry with characteristic mirror-symmetric "Collins-like" behaviour induced by the presence of the $H_{1q}^{\bot h}$ Collins FF. Corresponding theoretical curves are presented for comparison in Fig~\ref{fig:LSA_1h_X_th} together with HERMES experimental points~\cite{Airapetian:2005jc}.
%

Last genuine azimuthal LSA is the $A_{LL}^{\cos(\phiH)}$ term. The asymmetry appears to be small with no clear trend observed due to relatively large statistical uncertainties, see Fig.~\ref{fig:LSA_1h}. There are several model predictions available for this asymmetry, all predicting small (less than 4\%) effect. Within the statistical accuracy, COMPASS results are in agreement with theoretical predictions as shown in Fig.~\ref{fig:LSA_1h_X_th} (bottom left panel).

Last two azimuthal LSAs appearing in the SIDIS cross-section arise due to the presence of non-zero transverse component of the target polarization. Both $A_{UL^{\ell}}^{\sin(3\phiH)}$ and $A_{LL^{\ell}}^{\cos(2\phiH)}$ LSAs are found to be small and compatible with zero, see Figs.~\ref{fig:lN_LSA_1h},\ref{fig:LSA_1h_Mean}.

In this Letter the preliminary COMPASS results of SIDIS measurements of the longitudinal-spin-dependent asymmetries are presented for the first time. Compared to the similar studies published by the HERMES~\cite{Airapetian:2005jc} and CLAS~\cite{Avakian:2010ae} collaborations, COMPASS results are characterized by an unprecedented precision, covering much wider kinematic range. These new data are the best currently available input for modelling and theoretical studies of longitudinal spin structure of the nucleon.
%

\bibliography{DIS2017_L_biblio}{}

\providecommand{\href}[2]{#2}\begingroup\raggedright\begin{thebibliography}{10}

\bibitem{Perdekamp:2015vwa}
M.~Grosse~Perdekamp and F.~Yuan, \emph{{Transverse Spin Structure of the
  Nucleon}},
  \href{https://doi.org/10.1146/annurev-nucl-102014-021948}{\emph{Ann. Rev.
  Nucl. Part. Sci.} {\bfseries 65} (2015) 429--456},
  [\href{https://arxiv.org/abs/1510.06783}{{\ttfamily 1510.06783}}].

\bibitem{Boglione:2015zyc}
M.~Boglione and A.~Prokudin, \emph{{Phenomenology of Transverse Spin: Past,
  Present and Future}},  \href{https://arxiv.org/abs/1511.06924}{{\ttfamily
  1511.06924}}.

\bibitem{Aidala:2012mv}
C.~A. Aidala, S.~D. Bass, D.~Hasch and G.~K. Mallot, \emph{{The Spin Structure
  of the Nucleon}}, \href{https://doi.org/10.1103/RevModPhys.85.655}{\emph{Rev.
  Mod. Phys.} {\bfseries 85} (2013) 655--691},
  [\href{https://arxiv.org/abs/1209.2803}{{\ttfamily 1209.2803}}].

\bibitem{Kotzinian:1994dv}
A.~Kotzinian, \emph{{New quark distributions and semiinclusive
  electroproduction on the polarized nucleons}},
  \href{https://doi.org/10.1016/0550-3213(95)00098-D}{\emph{Nucl. Phys.}
  {\bfseries B441} (1995) 234--248},
  [\href{https://arxiv.org/abs/hep-ph/9412283}{{\ttfamily hep-ph/9412283}}].

\bibitem{Bacchetta:2006tn}
A.~Bacchetta, M.~Diehl, K.~Goeke, A.~Metz, P.~J. Mulders and M.~Schlegel,
  \emph{{Semi-inclusive deep inelastic scattering at small transverse
  momentum}}, \href{https://doi.org/10.1088/1126-6708/2007/02/093}{\emph{JHEP}
  {\bfseries 02} (2007) 093},
  [\href{https://arxiv.org/abs/hep-ph/0611265}{{\ttfamily hep-ph/0611265}}].

\bibitem{Collins:2011zzd}
J.~Collins, \emph{{Foundations of perturbative QCD}}.
\newblock Cambridge University Press, 2013.

\bibitem{Diehl:2005pc}
M.~Diehl and S.~Sapeta, \emph{{On the analysis of lepton scattering on
  longitudinally or transversely polarized protons}},
  \href{https://doi.org/10.1140/epjc/s2005-02242-9}{\emph{Eur. Phys. J.}
  {\bfseries C41} (2005) 515--533},
  [\href{https://arxiv.org/abs/hep-ph/0503023}{{\ttfamily hep-ph/0503023}}].

\bibitem{Lu:2014fva}
Z.~Lu, \emph{{Single-spin asymmetries in electroproduction of pions on the
  longitudinally polarized nucleon targets}},
  \href{https://doi.org/10.1103/PhysRevD.90.014037}{\emph{Phys. Rev.}
  {\bfseries D90} (2014) 014037},
  [\href{https://arxiv.org/abs/1404.4229}{{\ttfamily 1404.4229}}].

\bibitem{Wandzura:1977qf}
S.~Wandzura and F.~Wilczek, \emph{{Sum Rules for Spin Dependent
  Electroproduction: Test of Relativistic Constituent Quarks}},
  \href{https://doi.org/10.1016/0370-2693(77)90700-6}{\emph{Phys. Lett.}
  {\bfseries B72} (1977) 195--198}.

\bibitem{Anselmino:2006yc}
M.~Anselmino, A.~Efremov, A.~Kotzinian and B.~Parsamyan, \emph{{Transverse
  momentum dependence of the quark helicity distributions and the Cahn effect
  in double-spin asymmetry A(LL) in Semi Inclusive DIS}},
  \href{https://doi.org/10.1103/PhysRevD.74.074015}{\emph{Phys. Rev.}
  {\bfseries D74} (2006) 074015},
  [\href{https://arxiv.org/abs/hep-ph/0608048}{{\ttfamily hep-ph/0608048}}].

\bibitem{Airapetian:2005jc}
{\scshape HERMES} collaboration, A.~Airapetian et~al., \emph{{Subleading-twist
  effects in single-spin asymmetries in semi-inclusive deep-inelastic
  scattering on a longitudinally polarized hydrogen target}},
  \href{https://doi.org/10.1016/j.physletb.2005.06.067}{\emph{Phys. Lett.}
  {\bfseries B622} (2005) 14--22},
  [\href{https://arxiv.org/abs/hep-ex/0505042}{{\ttfamily hep-ex/0505042}}].

\bibitem{Avakian:2010ae}
{\scshape CLAS} collaboration, H.~Avakian et~al., \emph{{Measurement of Single
  and Double Spin Asymmetries in Deep Inelastic Pion Electroproduction with a
  Longitudinally Polarized Target}},
  \href{https://doi.org/10.1103/PhysRevLett.105.262002}{\emph{Phys. Rev. Lett.}
  {\bfseries 105} (2010) 262002},
  [\href{https://arxiv.org/abs/1003.4549}{{\ttfamily 1003.4549}}].

\bibitem{Adolph:2016vou}
{\scshape COMPASS} collaboration, C.~Adolph et~al., \emph{{Azimuthal
  asymmetries of charged hadrons produced in high-energy muon scattering off
  longitudinally polarised deuterons}},
  \href{https://arxiv.org/abs/1609.06062}{{\ttfamily 1609.06062}}.

\bibitem{Alekseev:2008aa}
{\scshape COMPASS collaboration} collaboration, M.~Alekseev et~al.,
  \emph{{Collins and Sivers asymmetries for pions and kaons in muon-deuteron
  DIS}}, \href{https://doi.org/10.1016/j.physletb.2009.01.060}{\emph{Phys.
  Lett.} {\bfseries B673} (2009) 127--135},
  [\href{https://arxiv.org/abs/0802.2160}{{\ttfamily 0802.2160}}].

\bibitem{Airapetian:2009ae}
{\scshape HERMES collaboration} collaboration, A.~Airapetian et~al.,
  \emph{{Observation of the Naive-T-odd Sivers Effect in Deep-Inelastic
  Scattering}},
  \href{https://doi.org/10.1103/PhysRevLett.103.152002}{\emph{Phys. Rev. Lett.}
  {\bfseries 103} (2009) 152002},
  [\href{https://arxiv.org/abs/0906.3918}{{\ttfamily 0906.3918}}].

\bibitem{Adolph:2012sp}
{\scshape COMPASS collaboration} collaboration, C.~Adolph et~al.,
  \emph{{Experimental investigation of transverse spin asymmetries in
  muon-proton SIDIS processes: Sivers asymmetries}},
  \href{https://doi.org/10.1016/j.physletb.2012.09.056}{\emph{Phys. Lett.}
  {\bfseries B717} (2012) 383--389},
  [\href{https://arxiv.org/abs/1205.5122}{{\ttfamily 1205.5122}}].

\bibitem{Adolph:2012sn}
{\scshape COMPASS collaboration} collaboration, C.~Adolph et~al.,
  \emph{{Experimental investigation of transverse spin asymmetries in muon-p
  SIDIS processes: Collins asymmetries}},
  \href{https://doi.org/10.1016/j.physletb.2012.09.055}{\emph{Phys. Lett.}
  {\bfseries B717} (2012) 376--382},
  [\href{https://arxiv.org/abs/1205.5121}{{\ttfamily 1205.5121}}].

\bibitem{Parsamyan:2014uda}
B.~Parsamyan, \emph{{New results on transverse spin asymmetries from COMPASS}},
  \href{https://doi.org/10.1051/epjconf/20158502019}{\emph{EPJ Web Conf.}
  {\bfseries 85} (2015) 02019},
  [\href{https://arxiv.org/abs/1411.1568}{{\ttfamily 1411.1568}}].

\bibitem{Avakian:2007mv}
H.~Avakian, A.~V. Efremov, K.~Goeke, A.~Metz, P.~Schweitzer and T.~Teckentrup,
  \emph{{Are there approximate relations among transverse momentum dependent
  distribution functions?}},
  \href{https://doi.org/10.1103/PhysRevD.77.014023}{\emph{Phys. Rev.}
  {\bfseries D77} (2008) 014023},
  [\href{https://arxiv.org/abs/0709.3253}{{\ttfamily 0709.3253}}].

\bibitem{Mao:2016hdi}
W.~Mao, X.~Wang, X.~Du, Z.~Lu and B.-Q. Ma, \emph{{On the cos $\phi_h$
  asymmetry in electroproduction of pions in double longitudinally polarized
  process}}, \href{https://doi.org/10.1016/j.nuclphysa.2015.10.004}{\emph{Nucl.
  Phys.} {\bfseries A945} (2016) 153--167}.

\end{thebibliography}\endgroup

\end{document}